\documentclass[twocolumn,showpacs,preprintnumbers,amsmath,amssymb]{revtex4}

\usepackage{epsf}

 \newcommand{\insertplot}[5]{\begin{figure}
 \hfill\hbox to 0.05in{\vbox to #5in{\vfill
 \inputplot{#1}{#4}{#5}}\hfill}
 \hfill\vspace{-.1in}
 \caption{#2}\label{#3}
 \end{figure}}
 \newcommand{\inputplot}[3]{
 \special{ps: plotfile #1}
}

\date{\today}

\newcommand{\la}{\lambda}

\newcommand{\f}{\phi}
\newcommand{\vf}{\varphi}
\newcommand{\F}{\Phi}

\newcommand{\si}{\sigma}

\newcommand{\ee}{\end{equation}}
\newcommand{\eea}{\end{eqnarray}}
\newcommand{\be}{\begin{equation}}
\newcommand{\bea}{\begin{eqnarray}}

\newcommand{\pa}{\partial}

\newcommand{\vep}{\varepsilon}

\newcommand{\re}[1]{(\ref{#1})}
\newcommand{\R}{{\rm I \hspace{-0.52ex} R}}

\begin{document}

\title{Non-Abelian Yang-Mills--Higgs vortices}
\author{{\large Francisco Navarro-L\'erida}$^{\ddagger}$
and {\large D. H. Tchrakian}$^{\star, \dagger}$ \\ 
$^{\ddagger}${\small Dept.de F\'isica At\'omica, Molecular y Nuclear, Universidad Complutense de Madrid, E-28040 Madrid,
Spain}\\ 
$^{\star}${\small School of Theoretical Physics, Dublin Institute for Advanced Studies, 10 Burlington
Road, Dublin 4, Ireland }\\
$^{\dagger}${\small Department of Computer Science,
National University of Ireland Maynooth, Maynooth, Ireland}}

\begin{abstract}
In this Letter we present new, genuinely non-Abelian vortex solutions in SU(2)
Yang-Mills--Higgs theory with only one {\it isovector} scalar field. These non-Abelian solutions
branch off their Abelian counterparts (Abrikosov--Nielsen-Olesen vortices)
for precise values of the Higgs potential coupling constant $\la$. For all values of $\la$, their
energies lie below those of the Abelian energy profiles, the latter being logarithmically divergent as $\la\to\infty$.
The non-Abelian branches plateau in the limit $\la\to\infty$ and their number
increases with $\la$, this number becoming infinite.
For each vorticity, the gaps between the plateauing energy levels become constant.
In this limit the non-Abelian vortices are non-interacting
and are described by the {\it self-dual} vortices of the $O(3)$ sigma
model. In the absence of a topological lower bound, we expect these non-Abelian vortices to be $sphalerons$.
\end{abstract}


\maketitle

{\sl Introduction}

Non-Abelian vortices on $\R^2$ have attracted interest since a very long time. Nambu~\cite{Gervais:1976zg} pointed out
that vortices of finite length in $\R^3$ require monopoles at each end. Originally, they were proposed
by Mandelstam~\cite{Gervais:1976zg} as flux tubes absorbed by non-Abelian ('t~Hooft-Polyakov) monopoles at each end.
In this picture the monopoles are bound, implying that in the dual picture where the duals of the monopoles are the
quarks, one can describe confinement in QCD.

The 't~Hooft-Polyakov monopole is a topologically stable and finite energy solution of the $SU(2)$ Yang-Mills--Higgs
(YMH) system on $\R^3$, where the Higgs field takes its values in the algebra, $i.e.$, that it is an $isovector$,
$\vec\f=(\f^1,\f^2,\f^3)$,
under $SO(3)$ rotations. Topologically stable and finite energy vortex solutions of the gauged Higgs system on $\R^2$ on
the other hand are supported by the Abelian Higgs model, where the Higgs field is a complex scalar,
$\vf=\f^1+i\f^2$, $i.e.$, it is an
$isovector$
$\f^M=(\f^1,\f^2)$ under $SO(2)$ rotations. This is the Abrikosov--Nielsen-Olesen (ANO) vortex~\cite{Nielsen:1973cs}.
The field multiplets in the two models do not match.

To construct a non-Abelian vortex on $\R^2$, it was realised by Nielsen and Olesen that it is necessary to have
a model with more than one Higgs field.
They chose~\cite{Nielsen:1973cs} two $SO(3)$ $isovector$ Higgs fields,
each with its own symmetry breaking potential and vacuum expectation value (VEV), but with the vacuum value of each
oriented at different directions in isospace -- in
the simplest case being orthogonal to each other. In this way the $SO(3)$ gauge group is completely broken on the
asymptotic circle of $\R^2$, which is necessary for topological stability. Subsequently this construction was
extended in models featuring $N$ distinct Higgs fields, generalising the $SU(2)$ vortices of
\cite{Nielsen:1973cs} to $SU(N)$ in
\cite{deVega:1976rt,deVega:1986hm,Suranyi:1999dm,Schaposnik:2000zt,Kneipp:2001tp,Konishi:2001cj,Marshakov:2002pc}.
These vortices, described as $Z_N$ vortices, are not
genuinely non-Abelian since their flux is restricted to a single direction along the Cartan subalgebra.

More recently, this problem was considered in the context of ${\cal N}=2$ supersymmetric QCD models by Hanay and
Tong~\cite{Hanany:2003hp} by Auzzi $et.\ al.$~\cite{Auzzi:2003fs} and by Eto $et.\ al.$~\cite{Eto:2005yh}.
The salient feature of these models is that they
have both gauge and colour symmetries that are broken by the condensate of the scalar fields in such a way that the
unbroken subgroup results in orientational zero modes of the string, responsible for non-Abelian flux.

Non-Abelian vortices have been studied intensively in the context of dual confinement in QCD (see the reviews
\cite{Shifman:2007ce} and \cite{Eto:2006pg}). In addition to this physical application, they present
important examples of cosmic strings~\cite{Kibble:1978vm,Hindmarsh:1994re}, relevant to cosmological phase
transitions.

In this Letter we have constructed non-Abelian vortices of a $SU(2)$ YMH model with only {\bf one} algebra valued, $i.e.$,
isovector, Higgs field. (Non-Abelian vortices in the Weinberg-Salam model were constructed in \cite{Volkov:2006ug}.)
This model features exactly the same field multiplets, on $\R^2$, as the YMH system supporting
the 't~Hooft-Polyakov monopole on $\R^3$, differing from the two-Higgs models of \cite{Nielsen:1973cs} and those~
supporting $Z_N$ vortices, and obviously from the SQCD models of \cite{Hanany:2003hp,Auzzi:2003fs}.

{\sl YMH model and non-Abelian Ansatz}

Our model on $\R^2$ is described by the static Hamiltonian
\be
\label{H}
{\cal H}=-\frac12\,\mbox{Tr}\,F_{ij}^2-\mbox{Tr}\,(D_{i}\F)^2
+(4\la)^2\,\mbox{Tr}\,\left(\frac{1}{4}\upsilon^2+\F^2\right)^2\,,
\ee
where $\F=-\frac{i}{2}\vec\f\cdot\vec\si$ is the antihermitian isovector Higgs field, and
$A_j=-\frac{i}{2}\vec A_j\cdot\vec\si$, $j$ labeling the coordinate on $\R^2$,
with $\vec\si=(\si^M,\si^3)$, the Pauli matrices. 
The gauge field is defined by $F_{\mu \nu} = \partial_\mu A_\nu -
\partial_\mu A_\nu + [A_\mu, A_\nu]$ and the gauge covariant derivative is
given by $D_\mu = \partial_\mu + [A_\mu, \cdot]$. 
Note that only $A_j$, the magnetic components of the $SU(2)$ connection $A_{\mu}=(A_0,A_j)$,
appear in Eq.~\re{H},  since in the absence of a
Chern-Simons term, the electric component of
the connection $A_0$ vanishes, by virtue of the non-Abelian Julia-Zee theorem \cite{Julia:1975ff,Spruck:2008bm}.

The energy of this model is not endowed with a topological lower bound and {\it a priori} we would not expect the
resulting vortices to be topologically stable.
But the question of stability is more subtle than this. The (genuinely) non-Abelian vortices we have constructed
numerically, present bifurcations from the corresponding Abelian profiles, on plots of their energies {\it vs.}
the Higgs self-interaction coupling constant $\la$. Remarkably, it turns out that each non-Abelian profile lies below
the corresponding Abelian profile for all values of $\la$ and hence cannot be expected to decay into the Abelian
vortex with higher energy than it has. Indeed, we show that in the $\la\to\infty$ limit these non-Abelian vortices
are described by the (stable) {\it self-dual} 'instantons'~\cite{Polyakov:1975yp} of the $O(3)$ sigma model on $\R^2$.
However, since these Abelian vortices embedded in the non-Abelian theory at hand are known to be unstable, one
cannot expect this feature of the non-Abelian vortices found here to imply stability. Furthermore, the
Belavin--Polyakov 'instantons' feature an arbitrary scale, which indicates instability. Thus, we would expect that
our non-Abelian vortices are in effect, $sphalerons$.
The quantitative stability anaysis will be carried out elsewhere.

The radial Ansatz we use is 
\bea
\F &=& \upsilon h \, \frac{\sigma^{(n)}_r}{2 i} - \upsilon g \,
\frac{\sigma^3}{2 i} \ , \label{ansatz1} \\
A_j &=& - \frac{(\vep \hat x)_j}{r} \left( c \, \frac{\sigma^{(n)}_r}{2 i} - (a+n)
\, \frac{\sigma^3}{2 i}\right) \ , \ \ 
j=1,2 \ , \ \ \ \label{ansatz3} \nonumber
\eea
where 
we denote $\sigma^{(n)}_r = \cos n\vf \, \sigma^1 + \sin n\vf \, \sigma^2$ and
$(\vep \hat x)_j = ( \sin \vf, -\cos\vf)$. Here $\{a,c,g,h\}$ are functions
of $r$ only and the integer $n$ is the vortex number. This Ansatz, previously
used 
to construct
non-Abelian Chern-Simons--Higgs vortices~ \cite{NavarroLerida:2008uj}, is a consistent truncation of the most general
Ansatz.

{\sl Equations of motion}

Subject to the Ansatz Eq.~\re{ansatz1}, the Euler-Lagrange equations reduce to the following set of
non-linear ordinary differential equations,
\bea
&&-r\,\left(\frac{a_r}{r}\right)_r=\
 -\upsilon^2\,(a\,h-c\,g)\,h \, ,\label{br} \nonumber\\
&&-r\,\left(\frac{c_r}{r}\right)_r=\upsilon^2\,(a\,h-c\,g)\,g \, ,\label{dr}\\
&&(r\,h_r)_r=\frac1r(ah-cg)\,a
- 8 \upsilon^2\,\la^2\,r\,[1-(h^2+g^2)]\,h \, , \label{hr} \nonumber\\
&&-(r\,g_r)_r=\frac1r(ah-cg)\,c
+ 8 \upsilon^2\,\la^2\,r\,[1-(h^2+g^2)]\,g   \, , \ \ \ \ \ \label{gr} \nonumber
\eea
together with the constraint equation 
\be
\label{trunc_constr}
\upsilon^2\,(h\,g_r-g\,h_r)
-\frac{1}{r^2}(a\,c_r-c\,a_r)=0\,.
\ee
 The subscript $r$ denotes ordinary differentiation with respect to $r$.
The energy density now reads
\bea
{\cal E} &=& \frac{1}{4r^2}(a_r^2+c_r^2)  
+\frac{1}{4}\upsilon^2\left[(h_r^2+g_r^2)+\frac{1}{r^2}(ah-cg)^2 \right.
+  \nonumber \\
&& \left. 4 \upsilon^2\,\la^2\,[1-(h^2+g^2)]^2\right] \label{energy_density} \, ,
\eea
the total energy $E$ being given by $E=2\pi\int r {\cal E} dr$.

The embedded Abelian solutions, namely the solutions to the embedded Abelian subsystem, correspond to
the truncation $\{c=0, g=0\}$. These are the
ANO vortices which play an important role in the
classification of the non-Abelian vortices we have constructed. In particular we will study the dependence of these
on the parameter $\la$, so it is pertinent at this point to note that the critical configuration of the Abelian
vortices corresponds to the value $\la=\frac14$. This is the Bogomol'nyi limit where the Abelian vortices do not interact.


\begin{figure}[t]
\begin{center}
\epsfysize=6.5cm
\mbox{\epsffile{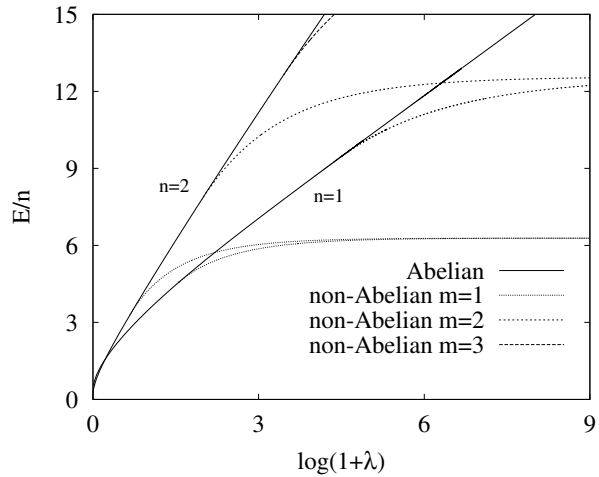}}
\caption{
Energy per vortex number $E/n$ versus the Higgs potential coupling
  constant $\la$ for YMH solutions with $n=1,2$.
}
\end{center}
\end{figure}

{\sl Numerical results}


In order to generate vortex solutions to Eqs.~\re{dr}-\re{trunc_constr}, we
impose boundary conditions such that the energy of the solutions is finite and
both gauge and the Higgs field functions are regular at the origin.
The system of equations is solved numerically
by means of a collocation method for boundary-value ordinary
differential equations, equipped with an adaptive mesh selection procedure.

The only free parameters are $n$ and $\la$, since we
fixed the unit of length by setting $\upsilon=1$ in what follows. For fixed
finite values of these parameters only a finite number of regular solutions
exist. There always exists one Abelian solution (ANO solution) for any
$n$ and $\la (\neq 0)$. For small values of $\la$ this is the only possible
solution. However, as $\la$ increases new non-Abelian solutions branch off the
Abelian ones. With increasing $\la$ more and more non-Abelian branches
appear, their number becoming infinite for $\la=\infty$. For given $n$ all the
non-Abelian solutions have energy lower than that of their Abelian
counterparts for each value of $\la$. This branch
structure of the solutions is exhibited in Fig.~1 where the energy per vortex
number, $E/n$, is plotted versus the
constant $\la$ for $n=1,2$. 

In this figure we observe the first two non-Abelian branches for $n=1$ and the
first three ones for $n=2$. The lowest non-Abelian branch branches off the
Abelian solutions at $\la \approx 3.705, 0.975$, for $n=1,2$,
respectively. It is clearly seen that higher values of the vorticity $n$
allow for new non-Abelian branches at lower values of $\la$. This is more
explicitly shown in Fig.~2, where the locations of the first branching points
for $n=1,2,3$ are given in logarithmic scale.

\begin{figure}[t]
\begin{center}
\epsfysize=6.5cm
\mbox{\epsffile{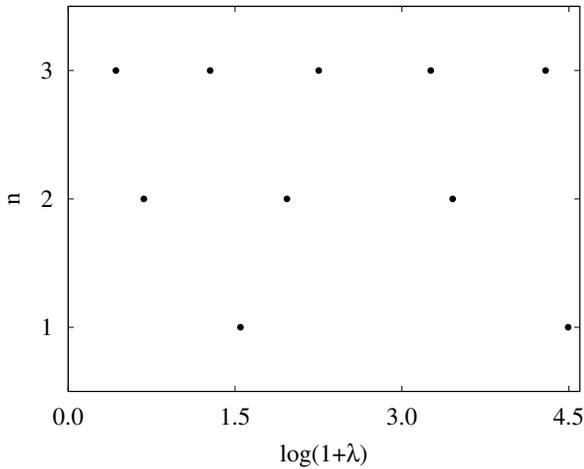}}
\caption{
Location of the first branching points for $n=1,2,3$.
}
\end{center}
\end{figure}

The structure of non-Abelian branches may be labeled by the pair
$(n,m)$,
$n$ being the vorticity
and the integer $m$ indicating the specific
non-Abelian branch for that vorticity $n$ (lower $m$ means lower energy).
Notice that the Abelian solutions behave in a different way, which we will
emphasize below. In fact, although Abelian solutions exist for any
non-vanishing value of $\la$, for each non-Abelian branch there exists a
minimal value of $\la$, $\la_{(n,m)}^{\rm min}$ (which depends on $n$ and $m$),
below which the non-Abelian branch ceases to exist. In
fact, at that minimal value the non-Abelian branch matches the corresponding
Abelian branch for that value of $n$, so $\la_{(n,m)}^{\rm min}$ corresponds
to the branching points where non-Abelian branches start to exist.

\begin{figure}[t]
\begin{center}
\epsfysize=6.5cm
\mbox{\epsffile{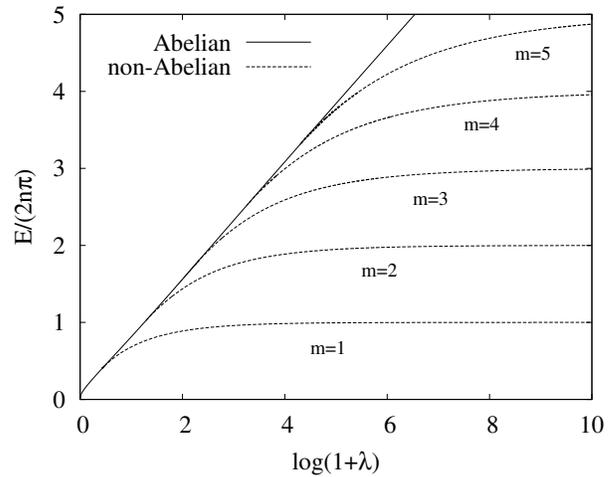}}
\caption{
Energy per vortex number $E/n$ versus the Higgs potential coupling
  constant $\la$ for YMH solutions with $n=3$. 
}
\end{center}
\end{figure}

A remarkable fact in Fig.~2 is that the gap between neighbouring branching points is
roughly constant for each $n$ on logarithmic scale for $\la$. More precisely, the quantity
$n \, [\log(1+ \la_{(n,m+1)}^{\rm min}) - \log(1+ \la_{(n,m)}^{\rm min})]$ is roughly constant and independent
of $(n,m)$. This feature becomes more accurate for large $\la$, revealing an
underying structure in the non-Abelian sector in the limit $\la \to \infty$. 
In fact, denoting the energy of the $(n,m)$ non-Abelian solutions by
$E_{(n,m)} = E_{(n,m)}(\la)$, one observes in Fig.~1 that for each  $m$ the
energy per vortex number tends to a limit which does not depend on $n$ but only on $m$.

It turns out that
\be
\lim_{\la \to \infty} \frac{E_{(n,m)}(\la)}{n} = 2 \pi m \,
\label{limit_energy}\,,
\ee
the energy {\it per unit} vorticity is equal to the energy of the {\it unit}
vortex. Hence non-Abelian vortices with given $m$ are non-interacting in that
limit. 
Fig.~3 shows this limit for $n=3$ solutions. In that figure it is clearly seen that in the large
$\la$ region the ratio $E/(2\pi n)$ approaches the integer value $m$ that labels the
non-Abelian branches. One observes an infinite number of non-Abelian branches for each $n$ emerging from the
logarithmically divergent Abelian profile, each converging to a finite limit.

One can understand this feature as follows. We have
verified that in this limit the contribution of the potential term in Eq.~\re{H} to the energy of the non-Abelian vortices
vanishes. (This contrasts with the corresponding situation for the vortices of the Abelian Higgs model.)
Thus, the YMH theory supporting the non-Abelian vortices becomes a $O(3)$ sigma model on $\R^2$ in this
limit~\cite{footnote}.
Likewise in our case the vanishing of the Higgs potential leads to
the $O(3)$ sigma model constraint, resulting in the $SO(3)$ gauged $O(3)$
model, which unlike in the WS case~\cite{Ambjorn:1984bb}, does not satisfy the
Derrick scaling requirement for finite energy. To this end, we have verified that  in this limit
the contribution to the energy of the YM term $\mbox{Tr}F^2$ in Eq.~\re{H} also vanishes, consistently with the Derrick
scaling requirement, and that indeed the YM potential becomes a {\it pure-gauge} in this limit. Thus the only
contribution comes from the $\mbox{Tr}D\F^2$ term in Eq.~\re{H}, which in this case reduces to $\mbox{Tr}\pa\F^2$
of the scale invariant $O(3)$ sigma model on $\R^2$. Our non-Abelian vortices in this limit are described by the
radially symmetric vorticity-$n$ subset of the
non-interacting {\it self-dual} Belavin--Polyakov ferromagnetic vortices~\cite{Polyakov:1975yp}.

\begin{figure}[t]
\begin{center}
\epsfysize=6.5cm
\mbox{\epsffile{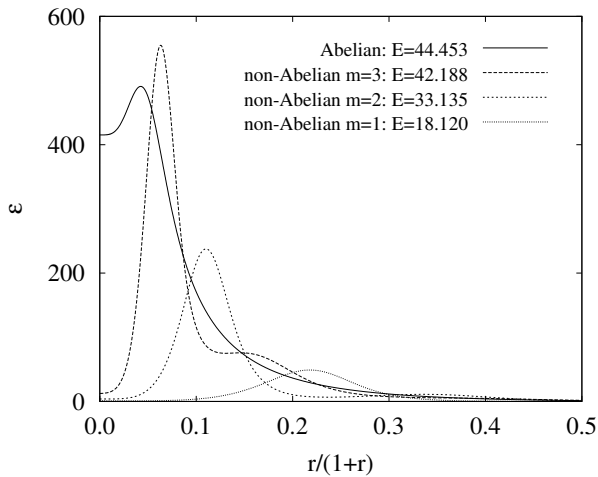}}
\caption{
Energy density for YMH solutions with $n=3$ and $\la=20.0$.
}
\end{center}
\end{figure}

The effect of non-Abelianness on YMH solutions affects not only the energy
values, which become lower for non-Abelian solutions, but also to the way the
energy is distributed throughout space. Both for
Abelian and non-Abelian configurations, solutions are radial for $n=1$
(their energy density having the global maximum at the origin) and circular
for $n > 1$ (their energy density having the global maximum at a finite
non-vanishing value of $r$). For ring-shaped configurations ($n>1$), as the
solutions are more non-Abelian (lower values of $m$) the energy density
profile spreads: the maximum is moved to higher values of $r$ and its height
decreases. In addition, the value of the energy density at the origin tends to
zero, the profile becoming more and more ring-like. This result is demonstrated
in Fig.~4 where the energy density profiles of YMH solutions with $n=3$ and
$\la=20.0$ are shown.

{\sl Conclusions}

As a final comment on the possible physical status of our solutions, we
emphasise that the model on $\R^2$ employed here is
precisely that which supports monopoles on $\R^3$. Interestingly, this
YMH model on $\R^3$ supports also monopole-antimonopole (MA) solutions, constructed in \cite{Kleihaus:2004is}.
This describes a consistent picture where our vortices are candidates for flux tubes
starting and ending on monopoles of opposite polarities. Our results are qualitatively consistent with the picture in
\cite{Kleihaus:2004is}.
In particular for vorticities $n\ge 3$, the energy density distribution in the MA configuration
presents a ring shaped density situated on the symmetry plane (the $\R^2$ plane where our vortices exist)
much like the circles in Fig.~4.

We conclude by noting that the vortices constructed are genuinely non-Abelian, but are not endowed with a
topological lower bound. That leads us to expect that these non-Abelian
solutions are unstable for finite values of $\lambda$, even though the
limiting solutions, namely the Belavin--Polyakov vortices for $\lambda \to \infty$ are stable.


{\sl Acknowledgement}

We thank Jurgen Burzlaff, Eugen Radu and Valery Rubakov for helpful discusions. This work is supported in part by
Science Foundation Ireland (SFI) project RFP07-330PHY, and, by
Minis\-terio de Ciencia e Innovaci\'on of Spain projects FIS2006-12783-C03-02, and FIS2009-10614.

\begin{small}

\end{small}

\end{document}